\documentclass{article}
\usepackage[utf8]{inputenc}

\usepackage{natbib}
\bibpunct{(}{)}{;}{a}{}{,}

\usepackage{slashed}

\usepackage{graphicx}
\usepackage{amssymb}
\usepackage{amsmath}

\usepackage{url}

\makeatletter
\newcommand{\manuallabel}[2]{\def\@currentlabel{#2}\label{#1}}
\makeatother

\manuallabel{ch:N-1}{6}
\manuallabel{ch:N-2}{7}
\manuallabel{ch:N-3}{8}
\manuallabel{ch:N-4}{9}

\manuallabel{sec:stringphys}{6.2}
\manuallabel{sec:StringQ}{6.3}

\manuallabel{fig:scat}{8.1}

\manuallabel{eq:closed}{6.6}
\manuallabel{eq:Ssig}{6.16}
\manuallabel{eq:ham}{6.20}
\manuallabel{eq:HamMass}{6.36}
\manuallabel{eq:open}{6.9}
\manuallabel{eq:sigmin}{6.21}

\title{Chapter \ref{ch:N-4}: The `emergence' of spacetime in string theory}

\author{Nick Huggett and Christian W\"uthrich\thanks{This is a chapter of the planned monograph \emph{Out of Nowhere: The Emergence of Spacetime in Quantum Theories of Gravity}, co-authored by Nick Huggett and Christian W\"uthrich and under contract with Oxford University Press. More information at www.beyondspacetime.net. The primary author of this chapter is Nick Huggett (huggett@uic.edu). This work was supported financially by the ACLS and the John Templeton Foundation (the views expressed are those of the authors not necessarily those of the sponsors). We want to thank Tushar Menon and James Read for exceptionally careful comments on a draft this chapter. We are also grateful to Niels Linnemann for some helpful feedback.}}

\begin{document}

\maketitle

\tableofcontents

\ 

This chapter builds on the results of the previous two to investigate the extent to which spacetime might be said to `emerge' in perturbative string theory. Our starting point is the string theoretic derivation of general relativity explained in depth in the previous chapter, and reviewed in \S\ref{sec:deriveGR} below (so that the philosophical conclusions of this chapter can be understood by those who are less concerned with formal detail, and so skip the previous one). The result is that the consistency of string theory requires that the `background' spacetime obeys the Einstein Field Equation (EFE) -- plus string theoretic corrections. But their derivation, while necessary, is not sufficient for spacetime emergence. So we will next, in \S\ref{sec:whenceST}, identify spacetime structures whose derivation would justify us in saying that a generally relativistic spacetime `emerges': this section will be important for establishing a fruitful way of approaching the question. The remainder of the chapter, \S\ref{sec:whencewhere}-\ref{sec:WFA}, will investigate how these structures arise as empirical phenomena in string theory: at this point we will also draw on the results concerning T-duality from chapter \ref{ch:N-2} (again summarizing the essential ideas). A critical question is the recurring one of this book: whether these structures are indeed emergent, or just features already present in the more fundamental string theory. Insofar as they are emergent, the goal is of course to try to understand the physical principles underwriting their formal derivation.



\section{Deriving general relativity}\label{sec:deriveGR}

So first, an overview of the derivation; refer to the previous chapter (which itself draws on \citet[\S3.4]{GreSch:87} and \citet[\S3.7]{Pol:03}) for more details. A caution: almost all of this section should be read as a discussion of a formal mathematical framework, rather than making metaphysical, ontological, or interpretational claims -- that will come in the remainder of the chapter. First a distinction.

We will use Feynman's `sum over paths' (or `path integral') formulation of quantum mechanics. The fundamental idea is that each \emph{classical} path $\gamma$ between states $A$ and $B$ (in a given time) is assigned a quantum amplitude $e^{iS(\gamma)/\hbar}$, where $S(\gamma)$ is the action along $\gamma$.  It's important to appreciate that \emph{every} path (for which $S$ is well-defined) connecting $A$ and $B$ is thus assigned an amplitude, not just the one that minimizes the action: every path counts, not only the one allowed by the classical equations of motion. Then the quantum probability amplitude for propagation from $A$ to $B$ is given by the \emph{sum} of path amplitudes: the amplitude for the system to evolve from $A$ to $B$ (in the given time) is $\sum_j e^{iS(\gamma_j)/\hbar}$, where the $\gamma_j$ are the different paths.\footnote{Of course, in typical cases the paths are not countable as suggested here, and the sum is replaced by an integral over paths -- hence the more common name for the approach. For current purposes, the difference is not important.} (The $A\to B$ transition probability is then obtained by squaring this amplitude.) The formulation can be applied to a particle to calculate the probability to propagate from one position to another: sum the amplitudes of every path between the two positions (and square). Or the probability that a field will evolve from one configuration to another: sum over all evolutions connecting them.

However, there is another famous sum associated with Feynman (building on the framework developed by Dyson and others), which should be carefully distinguished from that over paths: the sum over \emph{Feynman scattering diagrams}. An important aspect of Feynman's path integral formulation is that it makes approximation techniques largely mechanical in QFT: it yields a recipe for constructing successively smaller quantum corrections -- `perturbations' -- to a known classical system. `Feynman diagrams' are a powerful heuristic representation of these corrections, and summing them (if all goes well!) should yield successively better approximations to the probability amplitude. But there is nothing intrinsically perturbative about the path integral approach described above: in particular, \emph{a weighted path is not the same as a scattering diagram} -- in the case of fields, the former is a classical field configuration over spacetime, while the latter depicts a process involving (virtual) quanta. The sum over paths defines an exact quantum theory (specifically a complete set of amplitudes), while the sum over diagrams converges on the same set of quantum amplitudes. (At least, that situation is the  mathematical ideal: reality is typically less compliant.) But no proper part of one sum is the same as any part of the other. 

With these formal tools distinguished, we can review the derivation of GR (bearing in mind the caution that the following merely describes a formal framework, not an interpretation).\\

\noindent (A)  Feynman's perturbative approach is assumed to apply to quantum string theory. In fact the usual logic is reversed: instead of giving an exact formulation from which a perturbative theory can be derived by Feynman's methods, the theory is given in perturbative form, and an exact -- but currently imprecisely known -- theory which it approximates is postulated. (Most naturally one would suppose it to be a string field theory, e.g. \cite{Tay:09}, in which strings are created and annihilated, but most string theorists  believe that something more radical -- `M-theory' -- will be required.) Broadly speaking, the perturbative recipe is as follows. Suppose one is interested in the amplitude for a given group of incoming particles to interact and produce a given group of outgoing particles. Since (as discussed in chapter \ref{ch:N-1}) particles correspond to different excitations of strings, the given initial and final string states are fixed. Then the amplitude is computed through a perturbative sum over \emph{string scattering diagrams}.

\begin{figure}[htbp]
\begin{center}
\includegraphics[width=3.5in]{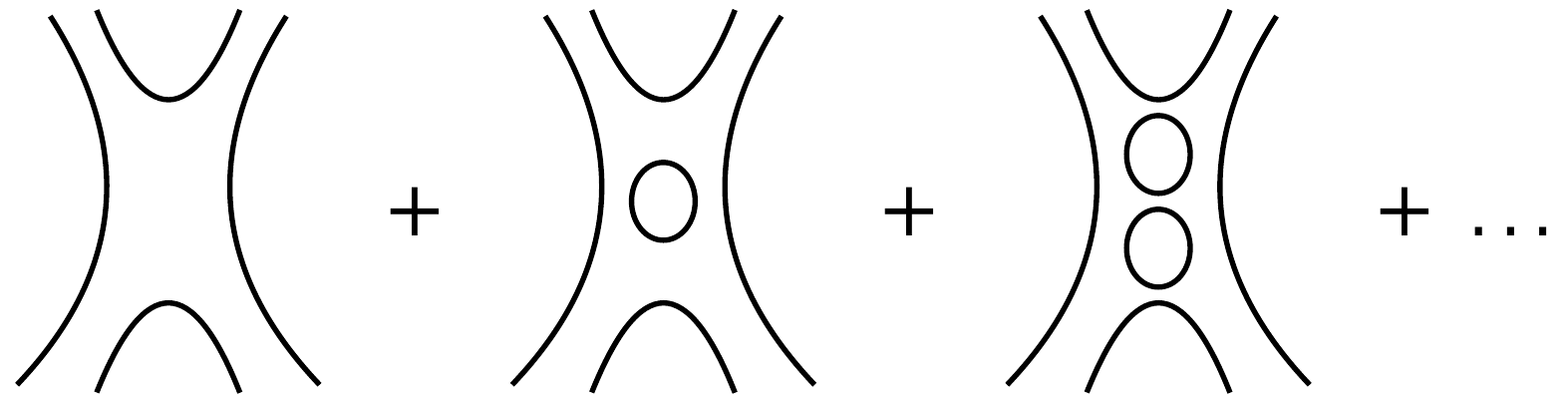}
\caption{The first three terms in the Feynman string path integral series: in the first, two strings join then split; in the second they join, then split and rejoin, before splitting; and so on.}
\label{fig:descryscatt}
\end{center}
\end{figure}

More specifically, each term in the sum corresponds to a different string worldsheet compatible with the given initial and final states: for example, two incoming strings might join to form one string, then split into two again; but they might also join, split, but then rejoin, and resplit; or they might split and rejoin twice; or three times; and so on (see figure \ref{fig:descryscatt}).  Each of these evolutions corresponds to a distinct worldsheet, with an increasing number of `holes'.%
\footnote{For the results discussed in this chapter the sum over topologies just described suffices. In fact, for full scattering calculations there are additional parameters -- `moduli' -- that need to be summed over in the path integral. These describe global properties of the worldsheet; for instance, how skewed a torus is (see chapter five of \cite{Pol:03}).}
 All need to be included in the sum: and the more splitting, the smaller the contribution from the diagram -- just as in particle scattering, in which a diagram contributes less the more interaction vertices it contains. (Because each interaction is associated with a `small' interaction factor $\lambda\ll1$, so that $n$ interactions suppress a term by a factor $\lambda^n$.)\\
 
\noindent (B) However, the probability amplitude for any given worldsheet will be the sum of the amplitudes of each of the different trajectories that the string might take through spacetime between given initial and final states. So the first perturbative sum over worldsheets contains a \emph{second} sum over all the different ways that each worldsheet might be embedded in spacetime: a non-perturbative (for now) sum over classical worldsheet paths, which we now explain.

To describe an embedding formally, as in chapter \ref{ch:N-1} we  assign spacetime coordinates $X^\mu$ to spacetime ($\mu=0,1,\dots D$, where there are $D$ spatial dimensions); and timelike and spacelike worldsheet  coordinates $\tau$ and $\sigma$, respectively, to the string itself, as a 2-dimensional spacetime object. Then $X^\mu(\tau,\sigma)$ is a function that takes each worldsheet point ($\tau,\sigma$) to the point of target space at which it is embedded, $X^\mu$. With flat metric $\eta_{\mu\nu}$, the Minkowski `sigma action' for a string with embedding $X^\mu(\tau,\sigma)$ is the worldsheet integral
    
  \begin{equation}
\label{eq:siggr}
S_\sigma[X^\mu] = -\frac{T}{2}\int\mathrm{d}\tau\mathrm{d}\sigma\ \eta_{\mu\nu}\big(\partial_\tau X^\mu\partial_\tau X^\nu - \partial_\sigma X^\mu\partial_\sigma X^\nu\big),
\footnote{See (\ref{eq:Ssig}). $\partial_\tau X^\mu$, e.g., is the spacetime vector expressing the rate at which the string's location changes with respect to its $\tau$ coordinate; $\eta_{\mu\nu}\partial_\tau X^\mu\partial_\tau X^\nu$ is its length squared. So the action is proportional to the difference between the squared rates of change with respect to $\tau$ and $\sigma$, integrated over the worldsheet.} 
\end{equation}
where $T$ is the constant internal tension. Thus, choosing units in which $\hbar=1$ (we also set $c=1$), the contribution to the amplitude from a particular worldsheet path is $e^{iS_\sigma[X^\mu]}$; and the total contribution of a particular worldsheet -- itself a \emph{summand} in the perturbative scattering series -- is the sum (or rather integral) of $e^{iS_\sigma[X^\mu]}$ over every possible target space embedding of the worldsheet, $X^\mu(\tau,\sigma)$. 
  
  Schematically then,
  
 \begin{equation}
\label{eq:sch1}
\textrm{total amplitude}\quad \approx \sum_{\textrm{worldsheets $j$}}\lambda_j\quad \cdot \sum_{\substack{\textrm{embeddings $k$} \\ \textrm{of worldsheet $j$}}} e^{iS_\sigma[X_{jk}^\mu]}.
\end{equation}  
 (\ref{eq:sch1}) represents the scattering dynamics of quantum string theory.\footnote{Amongst other things, a more rigorous treatment would Wick rotate the system into (complex) Euclidean space, $t\to it$.}  Reading from the right, each term $(j,k)$ corresponds to the $k$th embedding of the $j$th worldsheet, $X_{jk}^\mu(\tau,\sigma)$; their sum is an exact path integral for that worldsheet. But the first  sum, over $j$, is a perturbative Feynman  expansion over worldsheets of increasing number of holes; $\lambda_j$ decreases with $j$ because the contributions decrease the more string interactions occur. So overall the expression is a perturbative approximation. The first sum, over $j$, is often said to describe the quantum aspects of quantum strings, because it treats them much like field quanta; the second sum, over $k$, is correspondingly said to capture the peculiarly stringy nature of strings, because it depends on their extended nature, unlike QFT.  In the next two steps we will focus on this rightmost `stringy' sum, describing two of its crucial features.\\

\noindent (C) First, string excitations correspond to quanta, where the mass and type depends on the level of excitation: for example, a negative mass tachyon, or a massless graviton. (As discussed in chapter \ref{ch:N-1}, and below, these identifications make use of Wigner's idea that particle types can be distinguished according to the way their states transform under spacetime symmetries.) One can ask what difference it would make to its worldsheet amplitude if a string interacted with such a stringy particle during its evolution -- if, that is, a second, appropriately excited string interacted with the worldsheet. For reasons discussed in chapter \ref{ch:N-3}, the effect can be captured by including an appropriate factor in the path amplitude: specifically, the worldsheet integral of  a `vertex operator'. For instance, the vertex operator for a tachyon is $e^{ikx}$, and for a graviton $s_{\mu\nu}(\partial_\tau X^\mu\partial_\tau X^\nu - \partial_\sigma X^\mu\partial_\sigma X^\nu)$, with $s_{\mu\nu}$ a traceless symmetric tensor. That is, if the worldsheet interacts with a graviton, the appropriate path integral is now the sum over all embeddings $X^\mu(\tau,\sigma)$ of

\begin{equation}
\label{ }
e^{iS_\sigma[X^\mu]}\ \times \ \int\mathrm{d}\tau\mathrm{d}\sigma\ s_{\mu\nu}(\partial_\tau X^\mu\partial_\tau X^\nu - \partial_\sigma X^\mu\partial_\sigma X^\nu).
\end{equation}

If scattering occurs in the presence, not of a single quantum of gravity, but of $n$ gravitons, then the amplitude will have $n$ such factors. Note that exciting a single string doesn't produce more particles, but changes the type of particle associated with the string. Thus $n$ quanta means that there are $n$ similarly excited strings, and so a vertex operator for each one.


What though if scattering occurs in the presence of a classical gravitational field? In QFT, the appearance of classical fields is explained in terms of `coherent states', a superposition of every number of excited quanta: paradigmatically, a state of the form $\sum_n |n\rangle/n!$, where $|n\rangle$ is a state of $n$ quanta (this concept was explained in detail in an appendix to chapter \ref{ch:N-3}). So the path amplitude similarly involves a sum over powers of the graviton vertex operator

\begin{eqnarray}
\label{eq:CSPI}
\nonumber & & e^{iS_\sigma[X^\mu]}\ \times\ \sum_n \frac{1}{n!}\big(\int\mathrm{d}\tau\mathrm{d}\sigma\ s_{\mu\nu}(\partial_\tau X^\mu\partial_\tau X^\nu - \partial_\sigma X^\mu\partial_\sigma X^\nu)\big)^n\\
\nonumber & = & \exp\big(\int\mathrm{d}\tau\mathrm{d}\sigma\ \eta_{\mu\nu}(\partial_\tau X^\mu\partial_\tau X^\nu - \partial_\sigma X^\mu\partial_\sigma X^\nu)\big)\\
\nonumber & & \qquad \times\ \exp\big(\int\mathrm{d}\tau\mathrm{d}\sigma\ s_{\mu\nu}(\partial_\tau X^\mu\partial_\tau X^\nu - \partial_\sigma X^\mu\partial_\sigma X^\nu)\big)\\
  & = & \exp\big(\int\mathrm{d}\tau\mathrm{d}\sigma\ (\eta_{\mu\nu}+s_{\mu\nu})(\partial_\tau X^\mu\partial_\tau X^\nu - \partial_\sigma X^\mu\partial_\sigma X^\nu)\big),
\end{eqnarray}
where we substituted for $S_\sigma$ using (\ref{eq:siggr}), then used the Taylor expansion for $e^x$, and that $e^xe^y=e^{x+y}$.

But (\ref{eq:CSPI}) is exactly the amplitude we would have found if the Minkowski sigma action (\ref{eq:siggr}) had been modified by $\eta_{\mu\nu}\to g_{\mu\nu} = \eta_{\mu\nu}+s_{\mu\nu}$: the sigma action appropriate to a curved target space metric $g_{\mu\nu}$, rather than the flat one we supposed originally! In short, \emph{as far as the path integral is concerned, there is no difference between a curved spacetime, and a flat one with a coherent excitation of gravitons}. All this follows because the string action and the graviton vertex both involve a tensor coupled to $X^\mu$ derivatives (the latter because the tensor gravitational field $s_{\mu\nu}$ can be expanded in graviton plane waves, as discussed in chapter \ref{ch:N-3}); and because the quantum mechanical path integral exponentiates the action, while a coherent state exponentiates the graviton vertex operator. (There is nothing special about gravitons and the gravitational field in this analysis: similar points hold in the presence of coherent states of any of the fields composed of string quanta.)

To flag the crucial interpretational point that will arise from this mathematical fact: since all physical quantities can be derived from the path integral (according to the usual understanding), there simply is \emph{no} physical difference between the graviton and curved spacetime descriptions, and curvature \emph{is} a coherent state of gravitons. This idea is found in \cite[165-6]{GreSch:87} and \cite[108]{Pol:03} (amongst others), and with more conceptual detail in \cite{Mot:12}. That the metric is constituted by gravitons, rather than caused by them, and so of a different nature than in a classical spacetime theory, leads us to speak of its `emergence' (in our extended sense).\\

  \noindent (D) Second, now we change perspective in a significant way. A worldsheet has space and time coordinates, $\sigma$ and $\tau$ so formally is itself a (2-dimensional) spacetime.%
  \footnote{\label{ftnt:Weyl}As discussed in chapter \ref{ch:N-1}, these coordinates represent space and time in the sense that the worldsheet possesses a Lorentzian `auxiliary' metric $h_{ab}$,  chosen to be flat in (\ref{eq:siggr}). This should be distinguished from the `induced' metric inherited from the embedding in  spacetime. In particular, because of the `Weyl symmetry' (under $h_{\alpha\beta}(\tau,\sigma)\to \Omega^2(\tau,\sigma)h_{\alpha\beta}(\tau,\sigma)$) of string theory, $h_{\alpha\beta}$ does not encode metrical information beyond causal structure on the worldsheet (on which it agrees with the induced metric in classical solutions).}
  From this point of view, the value of $X^\mu(\tau,\sigma)$ at a point is formally the value of a $(D+1)$-component vector%
\footnote{Note that $X^\mu$ does not live in the tangent space to the worldsheet; from the point of view of the worldsheet $X^\mu$ lives in an `internal' vector space, on which $g_{\mu\nu}$ is the inner product.} at that point, rather than its spacetime location. From this perspective then, $X^\mu(\tau,\sigma)$ is not the spacetime embedding of the worldsheet, but a \emph{field} on the worldsheet, described by the action (\ref{eq:CSPI}); but these perspectives are formally equivalent, so the amplitude for a worldsheet embedding in spacetime is just the amplitude for the field $X^\mu(\tau,\sigma)$, which describes a classical evolution on the worldsheet. But now we realize that Dyson-Feynman perturbation theory can be applied to calculate an expansion for the sum over these classical field amplitudes: this is exactly the sum over embedding amplitudes in (\ref{eq:sch1}), which gives the contribution of each worldsheet.
  
Hence we can schematically rewrite (\ref{eq:sch1}) as
  
 \begin{equation}
\label{eq:sch2}
\left.\begin{array}{c}\textrm{total} \\\textrm{amplitude}\end{array}\right. \approx \sum_{\textrm{worldsheets $j$}}\lambda_j\quad \cdot \sum_{\substack{\textrm{$X^\mu$ field scattering} \\ \textrm{diagrams $k$ on} \\ \textrm{worldsheet $j$}}} (\alpha')^{n_k}G_{jk}.
\end{equation}  
$G_{jk}$ is the contribution of the $k$th scattering diagram for the $X^\mu$ field (picturing virtual processes for its quanta), evaluated on the $j$th worldsheet. $\alpha'\sim1/T$ is the `Regge slope', a small parameter in which the expansion can be made; its relation to the string tension gives another sense in which this sum is uniquely `stringy'. The perturbative sum is organized into powers of $\alpha'$, so into diminishing contributions. Compared to (\ref{eq:sch1}) we have replaced the exact path integral for $X^\mu$ with a perturbative expansion, so that now there is a doubly perturbative expansion. It is this second sum over $X^\mu$ scattering diagrams that is relevant to the derivation of the field equations of GR.

Thought of as describing a classical field theory on the worldsheet, $S_\sigma[X^\mu]$ is significantly different from familiar field dynamics in two ways. First, most fields depend on the metric of the spacetime in which they live, but $S_\sigma[X^\mu]$ does not depend on any worldsheet metric: it possesses `Weyl' symmetry. (A little more precisely, it is only dependent on the worldsheet lightcone structure, inherited from target space: the distinction between spacelike, timelike, and lightlike of curves restricted to the worldsheet. See footnote \ref{ftnt:Weyl} or chapter \ref{ch:N-1}.) As a matter of mathematical necessity, such a symmetry must also be possessed by the corresponding quantum theory; otherwise the resulting `anomaly' would render the theory inconsistent. 

Second, familiar fields have a constant interaction strength: for instance, electromagnetic interactions depend on the constant electrical charges of the quanta. But from (C) we know that in the presence of a gravitational field, or equivalently in a curved background spacetime, the action for the field on the worldsheet is given by (\ref{eq:siggr}) with $\eta_{\mu\nu}\to g_{\mu\nu}$. From the form of the action, it follows that when quanta of the $X^\mu$-field interact on the worldsheet, the strength of the interaction depends on $g_{\mu\nu}$,  which is a function of position on the worldsheet, not constant. Such quantum fields -- known as `sigma models' -- have been studied, and the critical result for us is that they are \emph{Weyl symmetric only if, at first order in $\alpha'$,  $g_{\mu\nu}$ satisfies the EFE}, which defines GR. Higher order terms produce corrections to the generally relativistic equations, but we emphasize that an expansion in $\alpha'$ is very different from a weak gravitational field approximation: it is the \emph{full} EFE that is derived, rather than its linear approximation.\footnote{We will say little more about the technical details of the derivation here, since they were described at length in the previous chapter (this chapter concerns the conceptual issues); but it is worth mentioning a couple of important points from that discussion. (i) If there are no other `background fields' (similarly understood as coherent states of string excitations), then the result is that $g_{\mu\nu}$ is `Ricci flat', the vacuum EFE. If there are other background fields, coherent states of other string quanta, then $g_{\mu\nu}$ will couple to them according to the corresponding EFE. (ii) The full analysis uses the stronger condition that the theory be Weyl symmetric at each order of perturbation theory. This condition provides an identity for $g_{\mu\nu}$ in powers of $\alpha'$. To first order in $\alpha'$, this identity is the EFE, while higher order terms produce small, stringy corrections to the classical solution.}
 \\
  
To run (C)-(D) backwards, we first learn (from Weyl symmetry) that quantum strings can only live in a curved spacetime that satisfies (to first order in $\alpha'$) the laws of GR. Then we find out that curved spacetime is nothing but a state of string graviton excitations, of strings themselves, in Minkowski spacetime. Together, those show that GR is an effective, phenomenal theory describing the collective dynamics of strings (in certain quantum states) in target space. The job of this chapter is to unpack this explanation, and try to understand better what is achieved, and how.\\

Before we proceed, it is worth pointing out the technical doubts that have been raised about the derivation of GR. If there are serious questions, is there any point considering what it tells us about the emergence of spacetime? In \emph{The Trouble with Physics},  (\cite[184-8]{Smo:07}) critically discusses the claim that the `derivation' described above amounts to a string theory prediction or explanation of gravity. His main points are that, first, the bosonic string theory which we have discussed is not a fully coherent physical theory because it contains a faster than light tachyon (which means that the theory is energetically unstable). Second, although this fact motivated string theorists to develop supersymmetric string theory to avoid the tachyon problem, such models are only known to exist in stationary spacetime backgrounds: for instance, black holes. They have not been shown to exist in evolving cosmological solutions that might describe our universe: for instance, in a Friedmann-Lema\^itre-Robertson-Walker spacetime. 
These points are valid restrictions on the argument we just reviewed: it doesn't \emph{prove} that a part of GR sufficient to describe our universe can be derived from supersymmetric string theory. Rather, the derivation within bosonic string theory, and the (as yet) partial success of supersymmetric string theory, are  \emph{evidence} that such a derivation is possible. 
 
After his largely negative assessment of this and other string theory `accomplishments', Smolin's conclusion is not that string theory should be abandoned, but rather that it should be one option pursued amongst others: as his work (and this book) exemplifies. In fact, string theory `succeeds at enough things so that it is reasonable to hope that parts of it, or perhaps something like it, might comprise some future theory' (198). (One possibility, of course, is that this future theory will be the `M-Theory' that string dualities putatively indicate. Then the derivation of the EFE, and the supersymmetric spacetimes are evidence that that theory can explain GR.) One could argue with his assessment, but because of the considerations of our introductory chapter, even Smolin's conclusion is sufficient to motivate our investigation of the `emergence' of curved spacetime in string theory. We explicitly expect to be operating in a field of incomplete models, to identify and understand the ways in which the spatiotemporal might be an effective form of the non- (or not fully) spatiotemporal, in a future more complete theory. Thus, while recognizing Smolin's important caveat, we will continue to analyze the question of spacetime emergence in the context of the bosonic string that we have described.

\section{Whence spacetime?}\label{sec:whenceST}

While the `emergence' of the metric from the graviton field is clearly crucial for the recovery of classical, relativistic spacetime (and of course for calculating quantum string corrections to GR), the discussion so far does not address how other aspects of classical spacetime arise in string theory. To what extent are they fundamental features of string theory? How do those aspects that are not emerge? As discussed in our Introduction, we follow a `spacetime functionalist' approach to make progress on these questions (see \cite{lamwut18,lamwut20}): we start with a list of relevant structures, which may or may not be exhaustive. Then we can attempt to trace out their origins in the theory. 

First a conceptual and terminological clarification. As we have stressed in earlier chapters, one should distinguish, at least conceptually, the classical spacetime of a theory like GR or QFT from the background, `target' spacetime of string theory. String theory stands in the relation of being more fundamental than, and (putatively) explanatory of GR or QFT etc; conversely GR and QFT are (putatively) effective descriptions of string theory in an appropriate limit. Although both fundamental and effective theories may involve formally similar structures -- in this case Lorentzian spacetimes -- one should not immediately conclude that these represent one and the same physical object; indeed chapter \ref{ch:N-2} argued that they do not. The general point is that formal models are representations of natural systems, and simply because two share a common formal structure, it does not follow that it represents a common natural object -- even if the models stand in the relation of fundamental and effective.

To mark the (at least conceptual) distinction between the spacetime of string theory and that of GR or QFT, it is useful to have a clear terminology (discussed further in the introduction). Thus we speak of the background spacetime assumed in string theory as `target space(time)', and the spacetime of GR or QFT as `classical spacetime', or `relativistic spacetime', but most often simply as `space(time)'. So, target space is (relatively) fundamental, while space may be effective or emergent. Or from a more epistemic point of view, spacetime is well-confirmed by the empirical success of QFT and GR, while target space is to be inferred from the success of string theory in recovering QFT and GR, and especially from successful novel predictions (were any to exist!).

Returning to the question of derived spacetime structures, suppose that one has a model of classical spacetime, compatible with available measurements: a smooth manifold, with a metric and matter fields, satisfying the relevant dynamical equations -- of GR, or of QFT, or of semi-classical gravity, depending on the situation described. Such a model has a topology, local and global, as well as a metric; we will enquire after the origin of each within string theory, focusing on the derivation of the EFE. 

Of particular salience, the local open set structure and the metric give meaning to locality in a general sense: the size of open regions, whether they overlap, how far apart they are, whether or not they are spacelike separated, and so on. Of course, the full open set structure of spacetime constitutes its global topology, but by a `local open set structure' we mean a proper subset of the open sets with their relations of overlap (subject only to the constraint that their union is a connected region). Informally, such a structure is the minimum required to make sense of `where' questions, such as `where did such-and-such an event occur?': a meaningful answer indicates an open set or sets (and excludes others). For our analysis, the paradigmatic events are \emph{scattering events}, so the paradigmatic answer will be that `the particles interacted inside \emph{that} region': the collision chamber of the CMS (compact muon solenoid) detector at CERN, say, a region of order 1m$^3$. Therefore, to understand the origin of the phenomenal local open set structure of  spacetime -- the location of particle scattering events -- we want to know how string theory represents such events; the answer could be the simple one that they are represented by the target space locations of stringy processes, but it might not be.
  
We want to make two points about the methodology just adopted. First, we are investigating the emergence of more than the EFE: by enquiring into the local open set structure, we are enquiring into the origin of the structure represented by the smooth manifold in a classical spacetime model. If you like, the derivation of the EFE just shows how classical gravity emerges in string theory, not how spacetime itself emerges. Second, by raising this question we are impinging on debates that go by the names `absolute-relative' or `substantivalism-antisubstantivalism' -- what does the manifold itself represent? On these questions readers may have strongly divergent presuppositions: manifold substantivalists (if any exist -- the term is from \cite{earnor:87a}) think that spacetime points are on an ontological par with physical entities, taking spacetime geometry quite literally; anti-substantivalists think not; others will simply not feel the pull of these debates at all. However, the question we have framed is neutral on such questions, because we are not asking about the origin of `the manifold itself', but about the derivation of the local open set structure of spacetime, which is something all should be able to interpret according to their preferred stance on classical spacetimes. We view this neutrality as an important virtue of the approach adopted here. (Of course, it may be that the analysis of the derivation itself bears on the issue of substantivalism, but we will not pursue that topic here.) 

To return to the question of emergence, thus framed, when one asks how a phenomenal theory -- one accommodating all current observations -- can be found as an `effective' description of a (more) fundamental theory in some regime, it is generally not a reasonable requirement that the phenomenal and fundamental theories be in perfect agreement. All that can be demanded is that they agree \emph{up to} the resolution of the (actual and possible) measurements that are taken to give empirical support to the phenomenal theory. After all, what we expect is that the fundamental theory will be a better match with experiment at some higher resolution: it will get things right that the phenomenal theory gets wrong.\footnote{Though it is also logically possible that the phenomenal theory gets some things more right than the more fundamental. Just because an account is better overall, it does not follow that it is perfect, or even better in every regard. That is, scientific progress need not be strictly cumulative.}
 
In particular, because the available energy places a practical restriction on the shortest measurable length, we should study spacetime emergence on the supposition of \emph{smallest measurable open sets} in the derived theory: in QFT or GR, especially. For instance, CERN's 14TeV Large Hadron Collider (LHC) probes at length scales around $10^{-19}$m. Any variations in quantities, including the metric, are indistinguishable from constant values across such sets:  a measurement of variation would amount, contrary to supposition, to a measurement within the set. We emphasize that such constraints on measurability and distinguishability are not a priori philosophical restrictions, but reflect the contingent limitations of the technology used to test the empirical consequences of the phenomenal theory. But the upshot is that (i) the local open set structure of the smallest measurable open sets, plus (ii) a spacetime metric compatible with them to experimental accuracy, plus (iii) the observable global topology, constitute empirical elements of classical spacetime which need to be understood within string theory. Our question thus is how these structures arise; whether they are in fact derived, or instead occur as fundamental components of string theory. To answer that, we need to examine the empirical basis of the structures.

In the derivation of GR from string theory, the phenomenal theory involves the scattering of quantum particles -- so QFT -- in a classical spacetime with metric $g_{\mu\nu}$, while string theory describes the scattering of corresponding strings in a target space of metric $g_{\mu\nu}$ (since quantum corrections to the metric are by assumption unobservable by the relevant measurements, we ignore them). (i) The open sets of interest are then those in which particle scattering events (real or possible) are located. (ii) In both string and field theory, $g_{\mu\nu}$ enters as an undetermined parameter in the action, and so appears as a variable in scattering cross-sections: thus with all other parameters fixed, $g_{\mu\nu}$ can be determined at the location of the scattering event, by the observed value of the cross-section.\footnote{A couple of important points should be made here, which will addressed later. First, note that instead of a more familiar appeal to rods and clocks, we are giving the metric `chronogeometric significance' through scattering cross-sections. Our choice has the added advantage that we do not introduce primitive objects external to the theory. Second, as we will discuss later, T-duality means that scattering does not in fact uniquely determine the background metric, as supposed here. This fact will not affect the arguments until it is addressed.}

At first glance these two spacetime structures seem to be directly given by corresponding structures of target space: it appears that (i) the measured open sets just are the locations of string scattering in target space, and (ii) the observed metric just is that of the string scattering cross-section. That is, target space just is the space of the phenomenal QFT, and our observations simply fail to resolve the spatial structure of strings, so that they appear as quanta. In the following sections we will argue that this correspondence is misleading (as of course should be expected  given (C) above and chapter \ref{ch:N-2}). 

We do not have a great deal to say about question (iii), of global topology. For as seen in chapter \ref{ch:N-2}, there are dualities which show that spacetime topology is not an invariant of string theory. Given our interpretation of dualities, it follows immediately that the observed topology is emergent not fundamental.\\

 A final methodological remark: we do not claim that the simple picture just sketched is a full account of how a stringy effective theory could be brought into practical, specific correspondence with a theory of classical spacetime. For example, \cite{Wil:06} investigates some of the many ways in which classical theories of different levels mesh, and convincingly shows that they never conform \emph{in detail} to the kind of story just sketched. And there is no reason to suppose that the situation won't be equally messy in the current case. But no matter. What is proposed is an idealized scheme to allow some points to be made about the relations between classical spacetime and string theory. To the extent that the idealization is accurate, the points are correct: such is always the situation, in physics or philosophy, when we wish to say something manageably succinct about a complicated reality. The only problem  (which Wilson calls `theory-T syndrome') arises if one mistakes such idealizations for claims about absolute metaphysical reality. But we won't do that! In particular, we don't claim that this construction is the most accurate possible description of our empirical knowledge of the topological and metrical properties of spacetime. The analysis only requires that it is good enough for our points to stand. To dispute that, it is not enough to show that our picture is incomplete: one would also need to show that our conclusions fail in a better picture. If that can be done clearly, we will indeed have a better understanding of these matters than that presented here.\\

\section{Whence \emph{where}?}\label{sec:whencewhere}

We first turn our attention to the origin of the observable open set structure of spacetime. As noted, things at first appear quite simple: naively, the open sets of space in which scattering events occur are nothing but the open sets of target space --  spatiotemporal open set structure is given at the fundamental level in string theory, and is in no sense `emergent'. But this appearance is deceptive, and we shall see that there are obstacles to this conclusion (\S\ref{subsec:dualscat}). Ultimately we will argue that localization is emergent, with the implication that strings are not literally spatial objects at all (\S\ref{subsec:scattop}). Before giving that argument we will explore an alternative to the target space view that seems to make the open set structure emergent; and show why it does not.

\subsection{The worldsheet interpretation}

One possible route to emergence was mentioned in \S\ref{sec:stringphys}: \cite{wit96} argues for a view that is popular, at least as a heuristic, amongst string theorists -- that the string \emph{worldsheet} rather than target space is the fundamental spacetime structure. We saw in (D) above that formally one can view the contribution of a given worldsheet as a scattering process for a vector field, $X^\mu$ living on the worldsheet: a picture in which the worldsheet is viewed as a 2-dimensional spacetime. Moreover, as discussed in chapter \ref{ch:N-1}, the 2-dimensionality of the string is important for the mathematics of the theory (its conformal symmetries, and the applicability of complex analysis). Witten's `worldsheet interpretation' proposes that this formal picture is in some sense the correct way to view perturbative string theory (until a more complete, non-perturbative formulation is available).

From this point of view, what is the origin of the open set structure of spacetime in GR or QFT? If the string worldsheet is the fundamental spacetime, then a first suggestion is that the origin lies in the open set structure of the worldsheet. This proposal (which Witten does not make) may not seem terribly plausible, since the worldsheet is only 2-dimensional. However, as described in chapter \ref{ch:N-2}, and discussed below, there are dualities that make spaces of different dimensions (at least) empirically indistinguishable. Moreover, the second series (over $k$) in (\ref{eq:sch2}) formally describes a quantum field on the worldsheet, so there is something to the idea, and it is worth disposing of it with a little care. The problem is that this proposal misinterprets the double sum over paths, for it is the first sum, over worldsheets $j$, that corresponds to \emph{particle} scattering. As we explained in (B), the particles in a QFT scattering process are represented in string theory by strings in given excited states: so the terms in a Feynman expansion for a quantum field correspond to the different string worldsheets. And we are taking the phenomenal open sets to be the locations of, paradigmatically, such particle scattering events, say some particular event occurring in the CMS detector at the LHC; hence to the worldsheets $j$ of the first sum (see figure \ref{fig:descryscatt}). Thus localization of observed scattering depends on the space in which the strings are localized, namely target space. The second series quantum mechanically describes the localization of a given world sheet in target space; and so presupposes the open set structure of target space. It is thus localization in target space that determines the `where' of the scattering event, and so it is to the open set structure of target space, not the worldsheet that we should look. Or rather, from the worldsheet perspective, we should look to the open set structure of the space of \emph{possible values} of the field, $X^\mu$, since that is how target space is interpreted.

So we seem to have returned to the original naive idea, which the worldsheet interpretation was attempting to replace: that strings are localized in classical, observed spacetime. But what about the fact that $X^\mu$ is treated as a \emph{quantum field}, rather than coordinates of a classical target space? Does that imply a sense in which relativistic spacetime arises from something other than a presupposed spacetime? No, because it is still the structure of the space of possible classical values of the $X^\mu$ field that do the work. From the target space perspective, the quantum string does not follow a single classical path in target space, but `explores' an extended region; the amplitude involves a sum over target space paths (\ref{eq:sch1}). From the worldsheet perspective, the quantum $X^\mu$ field does not have a single classical configuration over the string, but a range of configurations are explored; reflected formally in the same way in (\ref{eq:sch1}). But this sum over paths simply assumes the standard open set structure of target space/the manifold of possible values of $X^\mu$, and it is this structure that corresponds to the structure of the locations of phenomenal scattering events. In short, the worldsheet interpretation doesn't provide a different account of the origin of event locations from that in the straight-forward, naive view with which we started. Either way it is taken to be the open set structure of target space, whether this is interpreted as a spacetime or as a space of possible classical field values: either way, the phenomenal open set structure is simply given in string theory, not emergent.

Therefore, the only way to argue for the emergence of a spatiotemporal open set structure from the worldsheet would be to take a hard ontological line on Witten's interpretation: target space \emph{really} represents field values and not a spacetime. Then one could claim that spacetime emerges from something that is in fact a field, not itself a spacetime. It is unlikely that Witten himself intended such a strong claim, rather than intending to demonstrate the heuristic value and conceptual naturalness of the worldsheet point of view (his more forceful points regarding `the fate of spacetime', concern dualities). Nor do we think it especially fruitful to pursue arguments on the question of whether $X^\mu$ is `really' this or that, unless its relevance to the development of string theory can be established (for instance, perhaps one of the interpretations aligns better with whatever the degrees of freedom of the exact theory turn out to be). Indeed, we view it as a virtue of the way we have posed the question of the emergence of spacetime -- in terms of local open set structure -- that it has already led to a substantive question with some traction on the physics involved; while avoiding debating whether there `really' is a field or spacetime, a cousin of the substantivalism debate that we put to one side earlier. We have seen quite clearly that both the target space and worldsheet views say that the structure of phenomenal event locations is not emergent, but directly built into string theory.

\subsection{T-duality and scattering}\label{subsec:dualscat}

But not so fast! What we learned earlier of dualities should make us doubt such conclusions. We argued in chapter \ref{ch:N-2} that dual theories are physically equivalent, so that only features on which they agree are physical. But so far we have only considered how the observed open set structure arises in one theory; what about its duals? If it is derived from a different open set structure in different duals, then those structures are unphysical (beyond their `common core'), and so not assumed in string theory after all. They are in some sense mere formal representation, and the open set structure of spacetime must arise from some other string theoretic structure that \emph{is} invariant under dualities. Indeed, given the argument that target spacetime is not spacetime, it seems something like that must be the case. 

To develop this line of thought, for simplicity let's focus on T-duality, and the case of flat background spacetime.\footnote{We shall see later that T-duality can be extended to general background metrics, and of course we have discussed dualities that do more violence to the topology of spacetime. But the same basic considerations apply in such more complicated cases.} Recall, we claimed on the basis of a formal isomorphism that string theory on a space with a closed dimension of (large) radius $R$ is physically equivalent to a string theory in a space with (small) radius $1/R$ (in units in which the small `string length' constant $\ell_s=1$). We argued that as a result only those quantities on which the duals agree are physical. So for example, there is no determinate physical radius to target space; and since there is a determinate (effective) physical radius to the space of GR and QFT, it therefore cannot be target space. 

Of course, since T-dual target spaces have the same cylindrical topology, even with their reciprocal radii, they have exactly the same topological open set structures: that much is physical (with respect to T-duality). But this observation alone does not entail that the open set structure of space simply arises from a single shared topology. For the open sets of target space are mapped to those of space by identifying (possible) scattering events in each: again, as we discussed in the previous section, open sets are understood as possible answers to `where' questions about phenomenal processes, so understanding their origin requires identifying the corresponding fundamental stringy processes, and whatever stringy structure `localizes' them. But that cannot be achieved merely by considering the topology of target space; we need to identify the stringy process in each dual, to see what they have in common.

\begin{figure}[htbp]
\begin{center}
\includegraphics[width=3.5in]{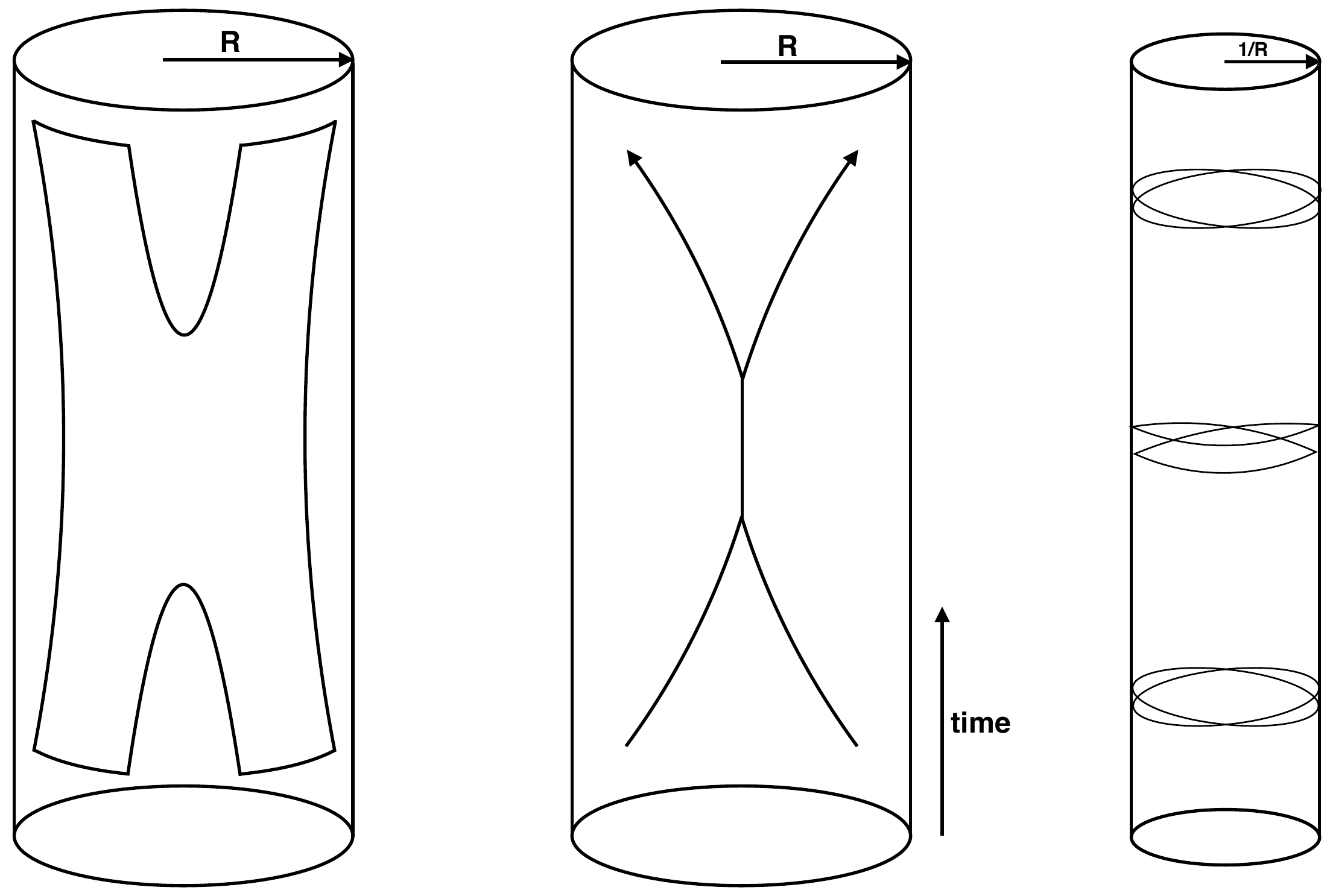}
\caption{Scattering in a space with a closed dimension (other spatial dimensions are not shown). In the center is the observed QFT process, in a dimension of radius R, in which two particles interact to produce a third, which then decays to two outgoing particles. On the left is the string description in a target space of radius $R$: two strings join into one, which then splits into two. On the right is the dual -- physically equivalent -- description in a target space of radius $1/R$: two once-wound strings join to form a single twice-wound string, which then splits back into two once-wound strings.}
\label{fig:dualscat}
\end{center}
\end{figure}

So consider a phenomenal particle scattering event: say two incoming quanta interacting at the LHC to produce two outgoing quanta. The radius, $R$, of the space in which we observe this event is of course large -- much larger than the CMS detector in which it occurs, say. This particle scattering event then has a string theoretic description in a large target space, also of radius $R$, in which the quanta are replaced by appropriately excited strings, in the way that we have discussed. The situation is pictured in figure \ref{fig:dualscat}, at the lowest `tree' level of perturbation theory, for an interaction allowing the production of an intermediate particle from the incoming pair. The key point, shown clearly in the figure, is that scattering occurs in a region of the cylinder homeomorphic to the plane, in space and in large radius target space.

But the kinetic energy of a string in a large radius theory is dual to the winding energy of the dual string: the potential energy it has due to being stretched around the closed dimension a number of times under its internal tension. So the T-dual scattering process is quite different: where before there were strings carrying kinetic energy, now there are wound strings. A process that involves strings joining and splitting in one dual corresponds to strings changing their winding numbers in the other dual. Thus the T-dual process is also pictured in figure \ref{fig:dualscat}, as two once-wound strings joining to form an intermediate twice-wound string, which splits into two again. Of course, these are dual -- physically equivalent -- representations of the single observed particle scattering process. Therefore, the T-dual region of target space in which scattering takes place is drastically changed; because the strings are wound, the whole of space is involved, a region with the topology of a cylinder. Compare this with the first dual, according to which scattering took place in a proper subregion, homeomorphic to the plane.\footnote{For now we will let the figure carry weight of the argument for this conclusion; below we will consider objections.}

Although the dual target spaces are homeomorphic cylinders then, the physical question of locality depends on how the locations of events in one space correspond to those in the other -- and T-duality does not preserve the topology of these locations, as this example shows. So the duals disagree over where in target space a scattering event occurs, and hence the representations of where scattering occurs in the duals are not fully physical (and only physical insofar as the duals agree): specifically, there is no physical fact about whether an event occurred in a planar open region, or in a cylindrical region of target space. Therefore, although the spatial location of a scattering event will formally agree with that of the large radius dual, since the latter isn't physical, phenomenal scattering regions are not directly built into string theory. On the other hand, just as for the phenomenal radius, there is a fact of the matter about the phenomenal scattering location: and so the locality structure of  space --  which one can say \emph{is} space in a pre-metrical sense -- is also derived or emergent in string theory, because of T-duality. 

Before we unpack this argument, we want to address the puzzle of how a process localized around an entire closed dimension of target space could possibly appear as if localized in an open region of space. In chapter \ref{ch:N-2} we discussed the similar question of how a target space of small radius appears as a  space with an observably large radius, and how \cite{BraVaf:89} answer. To recap: T-duality works because quantized strings require two spaces to represent states, target space, and a `winding space' of reciprocal radius; spatial wavefunctions in target space represent string momentum, while a `winding wavefunction' in winding space represents the quantum state of winding; under T-duality, spatial wavefunctions in target space are mapped to winding wavefunctions and vice versa; hence processes in a target space of radius $R$, are dual to processes in a winding space of equal radius, $1/(1/R)=R$; so if a phenomenal process that measures the radius of  space (timing a photon around the closed dimension is their example) corresponds to a process in a target space of radius $R$ in one theory, in its dual it corresponds to a process in a winding space of radius $R$ -- either way, the measurement will be $R$, an observed large radius.\footnote{The other assumption that is required to guarantee a \emph{large} radius is that the measurement is a low energy process: a long wavelength spatial wavefunction for the radius $R$ dual, and a low winding state for the the radius $1/R$ dual.}

The path signposted by Brandenberger and Vafa can be followed in this case too, to explain how a scattering event localized around an entire cylindrical dimension of a small radius target space appears as a scattering event localized in a planar open set of  space. The winding pictured in target space corresponds to a quantum wavefunction living in winding space: that's how the theory in fact represents winding. So the quantum description of scattering in the radius $1/R$ dual is as a process in winding space -- indeed, formally the very same processes found in target space in the radius $R$ dual. (So the leftmost, radius $R$ cylinder in figure \ref{fig:dualscat} can be interpreted as picturing a tree level winding space transition amplitude for its dual.)  That is to say, according to the small radius dual, the open set structure of  space comes from that of winding space instead of target space. And that's how the closed region of target space appears as an open region of  space.

\subsection{Scattering and local topology}\label{subsec:scattop}

With that puzzle resolved, we can turn to assessing the duality argument for the emergence of open set structure more carefully. We claimed that because the duals disagree on the localization of scattering in target space, such target space localization must be indeterminate; while the observed localization in classical space is determinate (setting aside ordinary quantum uncertainty). There were two steps to this argument. First, that the physical content of T-duals is restricted to their common part: whatever they disagree on is indeterminate. The argument for this claim was given at length in chapter \ref{ch:N-2}, so we will not rehearse it again here -- we claim that the move from metrical issues (the radius of space) to topological ones (the open set structure) does not change matters.

The second step of our argument notes that a scattering process occurs in an open region of target space  in one dual, and in a closed, cylindrical region in the other. Then the physical indeterminacy of target space localization follows immediately (as the indeterminacy of the radius did previously). And then because open set structure \emph{is} determinate for space, it cannot be identified with target space open set structure, but is emergent. However, so far the second step was supported simply by figure \ref{fig:dualscat}, picturing the dual scattering regions, which is rather quick.

That figure shows a tree diagram for a scattering process, rather than a full scattering event: only the first term of the leftmost sum in (\ref{eq:sch1}). What should we say of the full process, summed over all diagrams? Is that still localized in an open region? And even a single scattering diagram represents an integral over all of space, not just a proper subset of it: the rightmost sum in (\ref{eq:sch1}) embeds the worldsheet in every possible way in space, not just a subset of it. So is the diagram misleading about the localization of scattering? 

To answer these questions, consider that they are not specific to string theory, but apply equally to scattering in QFT, in which we certainly take concrete scattering events to be effectively localized, in the CMS detector, say. What holds in that case should also hold in the case of string scattering, which we take it to approximate. Indeed, (as spelled out in chapter \ref{ch:N-3}) the connection between string and QFT scattering diagrams is extremely tight. In the terms of the large radius dual, QFT scattering diagrams are taken to directly approximate string diagrams, at long length scales and low energies (relative to the string length and energy): see figure \ref{fig:scat}.\footnote{As previously noted, this identification has to be taken with a grain of salt: realistic, standard model particles do not arise simply as excitations of strings, but require specific tuning of background topology and a system of D-branes. The string diagrams that we are discussing do correspond to particle scattering diagrams, just (for the most part) not of particles observed in our universe. However, the argument pursued in this section still should go through in a more realistic scenario.} So if QFT scattering diagrams represent a localized process -- as indeed they do -- then so do the corresponding string diagrams.

Although a QFT S-matrix encodes amplitudes for particles coming from infinity to interact and produce new particles at infinity, integrated over the whole of space,\footnote{Note that the infinite volume approximation is not entirely apt: integrating over arbitrarily large distances can produce unphysical infrared divergences.} it is understood that this is a good approximation to the physical situation being modeled, in which the whole process is contained and observed in a finite region, say the CMS detector. Particle wavefunctions can never be localized for a finite time, so it is true that interactions of tails do happen over all space. But interactions are effectively localized, and of course for the identifications of phenomenal local regions, it is the effective localization that matters. Formally, that a finite volume interaction can be approximated by an infinite volume one is justified because in the limit that the finite volume goes to infinity, they agree (see \cite{Fra:06}, especially \S3.2.1, for a clear discussion). If the finite volume is `big enough', the differences will be unobservable. Then, since the particle diagrams are approximations to stringy diagrams, the latter can be taken in the same way, as approximating the same localized process, despite the sum over all target space. 

As for the small radius dual, the argument that the scattering region includes the entire circumference of the closed dimension is straight-forward: the in- and out-states are for winding eigenstates, and hence for strings that are wound around the dimension. There simply is no way for them to interact locally, unless the interaction is also around the dimension, for at least part of its duration. (And again, because T-duality exchanges space and winding space, represented in terms of the winding wavefunction, we have a processes effectively localized in winding space: formally the same process that occurs in target space in the large radius dual.)

And so indeed, the naive reading of figure \ref{fig:dualscat} is supported by more careful analysis: the duals disagree on how to represent the target space localization of a scattering event, and so target space is indeterminate with respect to that question. Moreover, the determinate localization of phenomenal scattering cannot simply be that of target space -- localization in classical spacetime is in that sense derived or emergent, not fundamental. 

Conversely, strings do not scatter in spacetime, taken as a literal fundamental statement: they scatter in target space (or, dually, winding space). As one describes nature at smaller and smaller scales, it is not the case that increasing spatial detail is included until particles are understood to be objects extended in one spatial dimension; instead, somewhere around the string scale the spatial picture breaks down altogether and is replaced by the distinct target/winding space one. In short, \emph{strings are not literally tiny objects}, because such a statement implies a comparison with spatial objects, which strings are not fundamentally. This is another aspect of the claim of string theorists that there is a shortest length scale, below which the concept of space breaks down.\\

That event localization is emergent in string theory is the main conclusion of this section, but it  serves another important purpose, independent of whether that result is accepted. What we have aimed to do with this example is illustrate the fruitfulness of the approach we have described for addressing the question of spacetime emergence; we specifically want to urge it as more fruitful, and better defined than attempting to import the substantivalism debate into quantum gravity. To summarize, identifying the locations of phenomenal events -- scattering events in this case -- has allowed us to ask a well-formed question about the origin of that location in the more fundamental theory: \emph{what is it about the fundamental description of the process that corresponds to that localization?} The answer might have been straight-forward: perturbative string theory is formulated in target space, so particle-event localization in space is simply understood as string-event localization in target space. But the conclusion of chapter \ref{ch:N-2} that target space is not space should already make that answer seem questionable. And indeed, the T-duality argument shows that there is no physical fact about the localization of string scattering in target space. Yet in either dual the string interaction is localized in some space (winding if not target), so the very notion of localization in an open set of \emph{some} manifold is still valid -- things are still `spatial' to that extent.

\section{Whence the metric?}\label{sec:whenceG}

The previous chapter, and the summary above, explained the standard account  of how a  metric satisfying the EFE can be derived in string theory as the excitations of graviton modes of strings: how string theory has GR, the classical theory of gravity, as a low energy limit. The key step in this derivation occurs in equation (\ref{eq:CSPI}). On the one hand, graviton coherent states (seen explicitly on the LHS of the equation) contribute to scattering exactly as a metric (seen on the RHS): they \emph{are} the gravitational field according to the theory. (And analogously for fields of other stringy quanta.) On the other hand, the observed, classical metric at a location (a smallest observable open set) is operationalized by the outcome of scattering in the location: in scattering cross-sections, $g_{\mu\nu}$ appears as a parameter, and so for fixed values of other parameters, can be measured by the observed cross-section. Specifically, a cross-section calculated using (\ref{eq:CSPI}) will measure $s_{\mu\nu}$, the tensor characterizing the graviton states, or equivalently $g_{\mu\nu}=\eta_{\mu\nu}+s_{\mu\nu}$, the observed metric.\footnote{As noted above, T-duality means that the fields are not uniquely determined. We will discuss this complication at the end of the section; for now we will ignore it for simplicity, as it will not affect the following arguments.} Weyl symmetry then entails that to first order the measured field satisfies the EFE; more sensitive measurements will, according to string theory, reveal stringy corrections. Or, from the graviton point of view, Weyl symmetry constrains the possible coherent states: $s_{\mu\nu}$ must be such that $\eta_{\mu\nu}+s_{\mu\nu}$ satisfies the EFE (plus higher order corrections). 

This account is central to the understanding of string theory as a theory of gravity, but it raises many questions, both to its conceptual and technical coherence, and to its claim to derive gravity as opposed to presupposing it. In this section and the next we will take up these issues, clarifying and defending the account as a derivation of gravity. This section deals with some of the broader conceptual issues; its sequel with some important technical limitations of the account.

\subsection{`Background independence'}

First, in a critical analysis, Penrose complains that ``string theory does not really properly come to terms with the problem of describing the dynamical degrees of freedom in the spacetime metric. The spacetime simply provides a fixed background, constrained in certain ways so as to allow the strings themselves to have full freedom." (\cite[897]{pen04}).\footnote{Chapter 31 of Penrose's \emph{Road to Reality} is highly recommended and insightful. It contains serious technical challenges to string theory as an exact, complete story, in opposition to the strongest claims for the theory. He does not deny that string theory is a promising partial story of quantum gravity, and so these points need not be addressed here. Once again, we assume only that string theory is at least a partial theory for the emergence of spacetime, not that it is the final truth: that is enough for the kind of work that we defended in the introduction.} This quotation alone may be a little ambiguous, but in the context of the chapter, the most reasonable reading is that Penrose is thinking of only one half of the derivation of GR: that avoiding the Weyl anomaly requires  the `background' metric to satisfy the EFE. His discussion of gravitons (896) appears to be restricted to the point that as excitations of the string they will contribute perturbatively to scattering processes. He does not mention the essential point (C) above, that in coherent states they contribute to the `background': that they \emph{are} the gravitational field. Only by considering the  partial picture that he presents does it appear that there is no more than a restriction on the kind of `container' in which strings can live. The full picture changes the situation completely.

To better understand this point, let us clarify an ambiguity in the term `background' (hence the scare quotes in the previous paragraph). Suppose we use (\ref{eq:CSPI}) to write the string action as $\int\mathrm{d}^2\sigma\ g_{\mu\nu}(\partial_\tau X^\mu\partial_\tau X^\nu - \partial_\sigma X^\mu\partial_\sigma X^\nu)$. \emph{Formally}, this action can be understood to describe a 2-dimensional hypersurface in a manifold with a given Lorentzian metric $g_{\mu\nu}$. The conventional terms for this metric and manifold are the `background metric field' and `background spacetime'. This terminology reflects the mathematical fact that formally one is studying quantum perturbations around a classical `background' solution, itself assumed to approximate some quantum state: a standard approach to quantum mechanics.\footnote{In quantum gravity (and general gauge theory) it was developed by DeWitt in the 1960s (\cite{DeW:67}; see also \citet[\S2.2]{Kie:12}). Concretely, his approach relies on the fact that S-matrix elements depend on asymptotic in and out states, which can be taken to be Minkowski to define particle states, while the interior, interaction region has an arbitrary classical geometry.} String theorists use these terms in this formal sense, which has no interpretational implication: its use is completely consistent with a physical understanding that fundamentally $g_{\mu\nu}$ describes a coherent state of strings, and only effectively a classical spacetime.

However, in the context of gravitational physics, and in Penrose's comment, `background' has a second meaning. GR is generally understood to make spacetime a dynamical object: in contrast with the given, fixed `background' of Newtonian or Minkowski spacetimes. This feature of GR is thus termed `background independence', and often urged as an important lesson of the theory, one that should be observed in a quantum theory of gravity (e.g., \cite{Rov:01} or \cite{Smo:06}). Thus, one might think that the `background' metric of string theory is a wrong turn!\footnote{Since the work of \cite{Kuh:62} philosophers and historians of science are more aware that such alleged deep principles are liable to change with major changes in theory. However, it seems at least reasonable to attempt a theory of quantum gravity that respects the principle: some restrictions on the possible avenues of exploration are needed.} 

But that of course would be to equivocate on the term: in the standard terminology of string theory the phrase `background metric' simply designates a part of the mathematical machinery, without any connotation that it is essentially non-dynamical. In particular, this use is not intended to express the claim that the theory is background dependent in the sense relevant to GR: that the metric is non-dynamical. Hence the question of whether  `background spacetime' (in the language of string theory) is `background independent' (in the language of GR) remains (as Rovelli and Smolin recognize). 

As we noted, the main part of Penrose's objection appears to be that string theory draws a distinction between space and its metric on the one hand (the `fixed background' to which he refers), and string dynamics, including graviton excitations, on the other. But if the discussion of coherent states is correct, then that view is incorrect: far from spacetime curvature being distinct from the string state, it is \emph{constituted} by it, and so could not be less distinct! As we have said before, the consistency requirement is on graviton states themselves, not on a distinct `container' in which the string dwells. In that sense the metric \emph{is} dynamical in string theory, because it captures the dynamics of the string, specifically of the graviton modes: a point made in \cite{HugVis:15} and \citet[2.4 -- chapter 6 gives a different, novel argument for the background independence of string theory]{Vis:19}.

This response to Penrose illustrates the difficulty in applying background independence as a desideratum in quantum theories of gravity. Attempts to formalize the notion in classical theories of gravity place conditions on the metric field: for example, that it be subject to the principle of least action (i.e., laws of motion) like other dynamical objects of the theory \citet[\S8 -- though as he points out, further conditions seem necessary]{Poo:15}. However, supposing that a classical criterion could be agreed upon, how is such a criterion to be applied to a quantum theory of gravity? The obvious approach of promoting it to a quantum condition may or may not be possible. If $g_{\mu\nu}$ is  promoted to an operator in some way, then one might attempt to `quantize' the criterion by, for example, stipulating that the metric operator field have a non-stationary Heisenberg picture dynamics. 

However, what if a theory has thoroughly non-spatio\-tem\-por\-al degrees of freedom? For example, we discussed the case of group field theory (GFT) in the introduction, whose degrees of freedom aren't spatiotemporal in all phases. And perhaps M-theory is the same: as we saw in chapter \ref{ch:N-2}, that is the lesson that \cite{BraVaf:89} and \cite{wit96} draw from T-duality. In such cases there need be no quantum `metric', whose dynamics can be subjected to a quantized criterion of background independence, so it is not clear how the question is to be posed. One might in this case think to apply one's criterion to the classical limit, but the motivation for so doing strikes us as weakened. Suppose the limit of an empirically successful theory failed some criterion: presumably this means that it only recovers a fragment GR, if full GR satisfies the criterion. But how would that be a strike against the theory? How could a theory with no fundamental spacetime at all be criticized for having a spacetime background in the sense of Newton or SR! One might better conclude that full GR is not the classical limit of fundamental physics, and that classical spacetime physics is not background independent.

The formal question of background independence is studied at much greater length in \cite{Rea:16} (see also \cite{Bel:11}), which reviews many different proposals, both classical and quantum, and gives a generally positive answer to the question of whether string theory is background independent in the various senses proposed. However, the question of background independence is secondary in the approach to emergence that we have outlined, which is to trace and analyze the \emph{derivation} of classical spacetime elements from the structures of string theory (and other quantum theories of gravity). Such an account is preliminary to determining to what extent string theory does or does not respect the `lessons' of GR; and  to judging how various formal notions might or might not capture those lessons in quantum gravity. (And indeed to understanding the origin of the dynamic nature of GR.) Once we have clarified what the fundamental structures of string theory are, and how classical spacetime is recovered, \emph{then} one can investigate how best to pose and evaluate the technical question of background independence. For present purposes, we are only interested in a looser notion of `background independence': to what extent are spacetime structures built in to string theory, and to what extent emergent, within the formal derivation we have described?\\

\subsection{Is there a Minkowski background?}

With the question of background independence acknowledged, and the somewhat different focus of our investigation into emergence emphasized, we should address an issue regarding metric emergence that will likely have occurred to the reader. Namely, even granted the graviton interpretation, in the action (\ref{eq:CSPI}), $g_{\mu\nu}$ is composed of two parts: $s_{\mu\nu}$ representing a tensor field contribution from the coherent graviton state, \emph{and} $\eta_{\mu\nu}$ representing a Minkowski metric. Doesn't that mean that a particular -- Minkowski -- metric structure is baked into string theory, rather than emergent? We will resist that conclusion, which will involve conceptually unpacking the derivation of the EFE more carefully. (Thereby also resisting a significant sense of background dependence: one invoked by \citet[21]{Ear:06}.) 

First, we note and put to one side three possible replies. (i) \citet[101f]{Rea:16} asks whether the Minkowski part of the background metric can also be interpreted as a coherent string state. He points out that $\eta_{\mu\nu}$ could never be a coherent state of gravitons, since it is not traceless (tr$[\eta_{\mu\nu}]=D-1$), while the graviton states are. But, he notes, there is a massless spin-2 string state with non-vanishing trace, composed of the graviton and a scalar excitation, which could form appropriate coherent states. (These states form a reducible representation of the spacetime symmetries, so correspond to composite particles in the usual understanding.) So could both parts of $g_{\mu\nu}$  be built as coherent states of stringy quanta in this way? He correctly replies that this idea won't work, because the very framework for talking about symmetries of string states, and the corresponding quanta -- not to mention the framework of QFT on which the notion of coherent state rests -- assumes Minkowski spacetime. (ii) If one followed Witten's worldsheet interpretation, then one would understand $\eta_{\mu\nu}$ to be the inner product on a vector field $X^\mu$, rather than a metric on `real' spacetime. But this response brings us back to the question of whether target space or the worldsheet is `really' the fundamental spacetime, which we have shelved. (iii) We note finally that string dualities mean that the background spacetime is not unique: T-duality means that the metric is not determinate, and duals with different topologies cannot be isometric. We will return to consider the implications of T-duality below, but our response to the issue of the Minkowski part of the metric does not appeal to duality. Instead, consider the following.

In arguing for the coherent state picture, we appealed to the fact that in equation (\ref{eq:CSPI}) $g_{\mu\nu}$ could be decomposed as $\eta_{\mu\nu}+s_{\mu\nu}$ without making any difference to the action, hence path integral, hence amplitudes. But the same holds trivially for an \emph{arbitrary} division of $g_{\mu\nu}$ into a metric and tensor field: $g_{\mu\nu} = \gamma_{\mu\nu}+s^\prime_{\mu\nu}$. (With $s'_{\mu\nu}= s_{\mu\nu} + \eta_{\mu\nu} - \gamma_{\mu\nu}$.) The split is irrelevant to the action, which only depends on the total, $g_{\mu\nu}$. As \cite{Mot:12} says, ``there's only one perturbative superstring theory in this sense -- whose spacetime fields may be divided to `background' and `excitations' in various ways.''\footnote{We have found many useful insights into string theory on his blog, although his posts are regrettably often marred by personal attacks. The piece cited (which addresses background independence) is free from these.}  If this is the case, the resultant metric is not just `emergent' in the sense that it involves a component from the string itself, but in the further sense that only the total has physical significance: there is no significance to any particular split into metric and coherent state contributions. In particular the metric part is conventional, without physical meaning. This argument, though we generally endorse it, requires some unpacking and qualifying, including the question of the significance of there being any split at all (\S\ref{subsec:whysplit}).

First some clarification of its terms: up to now, we have, in line with standard use, referred to $g_{\mu\nu}$ as the `background' metric, while in the passage quoted, Motl refers to its metric \emph{part} as the background! (Of course, because he is addressing the question of whether string theory is independent of this particular `background' structure.) For the purposes of our discussion of this proposal it will help keep things clear to adopt a different terminology: we will call $g_{\mu\nu}$ the `full' metric, and its components the `partial' metric and graviton/tensor field. Then we have argued that the full metric is emergent because of its graviton field part; but the question currently addressed is whether the partial metric is a non-emergent spacetime structure. The answer we take from Motl is that the partial metric is no structure at all, just an arbitrary component in a split of the only physical quantity, the emergent full metric. This split is necessitated, not by nature, but by the perturbative formalism currently used to describe string theory. (To make the existence of a formalism which does not require such a split a little more than a promissory note, he points out that string field theory, mentioned above, can be given without it.)

Then we reason as follows. The perturbative approach to quantum mechanics means selecting a classical solution with full metric $g_{\mu\nu}$, understood to approximately represent some quantum state, and then expand in quantum mechanical corrections. In just this way, perturbative string theory takes a solution to GR, and expands around it: along the way one finds that the background actually \emph{had} to be a model of GR. Moreover, the basic physical quantities that are defined within the theory -- scattering amplitudes (collectively, the `S-matrix') -- are identical to that of a model of involving stringy graviton excitations in Minkowski spacetime. That is, one could have taken that different classical solution as the starting point, and found the very same string physics, but within a different perturbative framework. But similarly, one could in principle have started with any (of some range of) classical solutions, with partial metric $\gamma_{\mu\nu}$, and studied quantum string perturbations around that: the excitations in that framework would also form collective states, including that represented by $s'_{\mu\nu}$. But since the path integral only cares about the full metric $g_{\mu\nu}$, once again the same S-matrix will be found.
 Although the formalism of perturbation theory requires a classical background, which one is an arbitrary choice, without direct physical significance.\footnote{It is worth pointing out that the S-matrix contains only asymptotic amplitudes, not the full set of correlation functions required to define an exact QFT via the Wightman axioms. Hence this equivalence is indeed firmly at the level of perturbative string theory -- which of course is more-or-less all that exists at present.} All that can be said is that the \emph{choice} of a Minkowski metric produces a simple and clear formal structure.

That is, as discussed in chapter \ref{ch:N-1}, in relativistic quantum mechanics -- including QFT -- elementary particle species are defined in terms of their (irreducible, unitary) representations of the symmetries of spacetime: how the particle states transform in Hilbert space under the action of the corresponding spacetime transformations.\footnote{This conception is due to \cite{Wig:39}, and is a staple of texts on relativistic QM and QFT: a particularly good treatment is found in \citet[chapter 2.5]{Wei:05}. Note that talk of `particles' in this context is loose (for instance, \cite{Mal:96}), and we do not intend to beg the question against such arguments concerning the particle concept in relativistic QFT. `Quanta' would be a better term, though see \cite{Rue:11} for an extended analysis of the tenability of even the standard quanta concept, in contexts relevant to curved spacetime. At base, our `particle' concept here just is Wigner's, of irreducible, unitary representation.} Spacetimes with different symmetries will have different (perhaps overlapping) representations, and hence may allow the existence of different kinds of particles, or no differentiation into particle species at all, if there are insufficient symmetries. However, the relatively small regions in which QFT is typically explored on Earth are suitably flat, so that the states of observable particles do form representations of the Poincar\'e symmetries. Thus, the identification of string states with particles that we have followed is based on their common representations of the Poincar\'e symmetries, and so on the assumption of Minkowski spacetime. That is, the existence of stringy quanta -- the scalars, tensors, etc that make up the string spectrum -- has, so far, been predicated on Minkowski geometry.

Thus, one cannot strictly describe the $s'_{\mu\nu}$ tensor field in an arbitrary metric+field decomposition as a `\emph{graviton} coherent state', for gravitons will not be defined in an arbitrary spacetime geometry. However, that is not to say that a similar conception is impossible in curved spacetimes: one needs to know, (i) on one side what form elementary particle states take, and (ii) on the other the string excitation spectrum, and see whether they can be identified. Away from flat spacetime, the formal situation is much less tractable, but in the next simplest cases, (i) there is a massless spin-2 graviton representation of the de Sitter and anti-de Sitter spacetime symmetries.\footnote{The issue of how to define mass is subtle: see \cite{Gar:03} for a recent treatment. The earliest classifications are found in \cite{Tho:41,New:50}.} While (ii) \cite{LarSan:95} find that there is indeed a massless spin-2 representation in the (closed) string spectrum which can be identified as the graviton, just as in Minkowski spacetime. So, indeed a split between partial metric and graviton field is not fixed spacetime structure, but can be chosen at least between Minkowski, de Sitter, and anti-de Sitter metrics. All the same, these particle identifications depend on a rich set of symmetries: in a general spacetime, Wigner's concept of particle will only hold approximately, in local patches with approximately symmetric metrics. In other backgrounds, the perturbative stringy excitations need not be `particles' in Wigner's sense at all; except approximately in small enough regions, or `at infinity' if spacetime is asymptotically Minkowski (say, if one considers scattering through a finite curved region). This fact does not change the arbitrary nature of the metric-tensor field split, but only shows that the string-quanta identifications are themselves an artifact of a particular approximation scheme. (This point will be addressed again, from a different perspective, in \S\ref{sec:WFA}.) But this itself is to be expected in a theory of quantum gravity in which the metric is not fundamental, for there is then no particular metric to which particle classification could be referred.

With these qualifications, however, we agree with Motl's proposal: string theory does not postulate a Minkowski background spacetime, in the sense that the choice of the Minkowski background in the equations is conventional. Other choices are logically equivalent, and the choice is made for mathematical convenience. In that case, no part of the full metric can be identified as `given', as it could equally (ontologically speaking) be redefined as a dynamical string state; and so we conclude that the metric is indeed derived not assumed in string theory.

\subsection{Why split the full metric?}\label{subsec:whysplit}

However, one might question the significance of splitting the full metric into tensor and metric field parts; after all, formally one could make the same split in classical gravity, without drawing any ontological conclusions. Indeed, that is how one approaches weak field GR, with a symmetric tensor gravitational field in Minkowski spacetime, described by the linear Fierz-Pauli action (e.g., \cite[chapter 18]{MisTho:73}). Of course the weak field approximation is not equivalent to GR: for instance, the point mass solution (the linear analogue of the Schwarzschild solution) implies 4/3 of the anomalous perihelion shift of Mercury (Misner et al., 183ff). There is no contribution of the tensor field to the energy-momentum tensor, and when it is included one is led to full GR.  But can't one still consider making a conceptual split between partial metric and tensor field, in solutions to full GR, with arbitrary choice of partial metric? That is, take a solution with metric field $g_{\mu\nu}$, formally split it into $\gamma_{\mu\nu}+s_{\mu\nu}$, and interpret the two terms as metric and gravitational tensor, respectively. Unless one was making such a split as a prelude to a weak field solution, this move seems arbitrary and unmotivated, but why precisely? And if the split is arbitrary and unmotivated in GR, is it also in string theory? In that case, one should not be splitting the total metric at all, so that it has no significance as a stringy graviton state, and is instead a given background -- and so assumed at the fundamental level, not derived or emergent.

Related questions have been explored from a different point of view in a recent paper (\cite{Rea:19}), on which we will base our argument for a principled difference between string theory and GR regarding the split. Read addresses the question of two `miracles' of GR (\cite{ReaBro:18}): these are a pair of related, arguably contingent facts about GR. (Since mere contingency is at stake, `miracles' may seem a little strong -- perhaps `maybes' would be more appropriate.) The first, which he designates MR1, is that the matter fields are all locally Lorentz invariant (and, one might add, take their standard forms in the same local coordinates). The second, MR2, is that the metric field has the same local symmetry as the matter fields. (Obviously, the latter follows from MR1 plus the local Lorentz symmetry of the metric.) The sense in which these two are contingent can be seen by considering a theory in which the metric is Lorentzian, but the matter fields are not: the EFE could still hold, because matter enters through the energy-momentum tensor, which does not determine the equations of motion for matter, or their symmetries. The drastic consequence of such a situation is that the matter fields would no longer reliably `measure' the metric field, because their transformations would no longer correspond to those of the metric: if a material measuring system in one frame were locally calibrated to space and time, a boosted duplicate would not be. So the two miracles/maybes are important to give the metric observable `chronogeometric' significance.\footnote{\cite{Rea:19} has rods and clocks in mind as the measuring systems, but the same point holds in our approach, in which the metric is `measured' by scattering amplitudes; if the S-matrix does not transform as the metric, in what frame are we supposed to take it?}

In string theory, however, both follow from the dynamics because of the underlying Minkowski metric: the $\eta_{\mu\nu}$ in either (\ref{eq:siggr}) or (\ref{eq:CSPI}).\footnote{Although we just argued that this particular split is arbitrary, it will be convenient to make this choice to discuss the question of a classical split. We don't see that the underlying point depends on this choice.} It means that all stringy quanta, whether matter or graviton, transform according to the same underlying Lorentz symmetries, entailing that their fields all have the same spacetime symmetries: of course, what is crucial here is that string theory \emph{unifies} matter and gravitons, as different string states. As Read concludes, unlike the case in GR, MR1-2 -- and hence the chronogeometric significance of $g_{\mu\nu}$ -- are not `miracles' (or even `maybes') in string theory, but a matter of physical necessity.\footnote{While the chronogeometric significance of $g_{\mu\nu}$ is necessary in string theory, Read agrees that it is still `emergent' in the sense that it is not a \emph{fundamental} element of the theory, exactly because it contains a stringy component.} 

This line of thought emphasizes the important difference between GR and string theory, and brings us back to the question of why splitting the metric into partial metric and tensor/graviton field parts is ad hoc in the former but not the latter. If a classical split is made, MR1 is unchanged, but MR2 then contingently asserts that the gravitational tensor field transforms as the matter fields. But there is nothing contingent about this fact in string theory, since the tensor and matter fields are on the same footing, as different string states. The contingency in GR emphasizes that nothing requires unification of metric and matter fields, and so splitting off an arbitrary tensor part of the metric does not pick out a new field on a par with the matter fields. On the contrary, in string theory the graviton field part of the split $g_{\mu\nu}$ is of a kind with matter fields, since both are string states. And so in that sense a classical split is unmotivated, but given -- though arbitrary -- in string theory. 

That essentially concludes our argument that the standard account of the metric in string theory indeed shows it to be derived, or emergent, rather than fundamental. Unlike a classical gravity tensor field, the stringy tensor field is of a kind with all other fields, composed of stringy quanta; moreover, no particular part of the metric field is a background rather than stringy. That said, some important issues remain; first, how does T-duality bear on this picture? And then some more technical questions about the validity and significance of the derivation.

\subsection{T-duality}

\cite{Rea:19} also raises the question of T-duality in the emergence of the metric; an issue discussed further in \cite{ReaMen:19}. A basic question is whether the duality is with respect to the partial metric ($\eta_{\mu\nu}$, say) or full background metric $g_{\mu\nu}$? The former we have already argued to be conventional. But if T-duality applies to the latter, then from the arguments of chapter \ref{ch:N-2} it will \emph{not} be identified with the observed, classical metric: that will have to be understood in terms of invariant measurements, along the lines described by Brandenberger and Vafa. To answer this question we turn to the `Buscher Rules', which describe how to implement T-duality in certain curved spacetime backgrounds.\footnote{\citet[\S14.2]{BluLus:12} is a good introduction} In particular, they describe the duality that occurs in the \emph{full} background $g_{\mu\nu}$, in the case of closed dimensions, with suitable symmetries of the metric: so in the terminology of chapter \ref{ch:N-3}, the full background metric is the `target space' metric. In those cases then, observations cannot uniquely determine the background metric: and, as in the simpler Minkowski case, low energy observations will agree on the metric observed, in both duals. In a Minkowski background we saw, for instance, that an experiment timing a photon around the universe would have the same result whether target space were large or small: a large observed radius. Read and Menon propose that the result would be the same in the more (but not completely) general case of the curved background T-duality: the large spacetime that we in fact do observe. We concur. Therefore, we need to revise our earlier understanding of how the metric (and other fields) are operationalized through scattering observations: scattering cross-sections are not fully observable, but only up to differences that would allow discrimination of dual metrics.\footnote{Note that scattering assumes in- and out-states at infinity, and so is not defined in a compact dimension. Local scattering measurements can measure the local metric, but scattering is not an appropriate way to measure global vertical quantities, such as the circumference. To do that, one needs an experiment like Brandenberger and Vafa's photon circumnavigation.}\\


In this section we have defended the claim that the metric field should be understood as derived in string theory against the major objections to this view. In so doing we have aimed to clarify the content of the claim, and especially its relation to the perturbative nature of string theory as it exists. However, while we endorse this position, we also argue that it is circumscribed, by its perturbative nature, and by the assumption that coherent states provide a classical limit of QFT, as the next section explains.

\section{Quantum field theoretic considerations}\label{sec:WFA}

The string theoretic interpretation of the metric works on the understanding (i) that suitable string states correspond to quanta of the gravitational field, and (ii) that such stringy gravitons form collective -- i.e., `coherent' -- states; and on the assumption (iii) that a classical metric field corresponds to such coherent quantum states, in the classical limit. We have discussed the reasons for holding (i) and (ii), but clearly (iii) is equally important, and relatively neglected. In this section we will therefore investigate this question, and argue that while generally justified, its scope is limited, and indeed importantly unknown in the case of gravity. The following discussion is necessarily more technical than the foregoing, and could be skipped by those wanting to understand the overall argument for the derivation of the metric, and read by those wanting a firmer grasp of the theoretical underpinnings of string theory. However, we hope that it  will stimulate further investigation of these important, but challenging questions.

\subsection{The graviton concept}

First, we should recognize that the concept of a `graviton' is dependent, not only on  spacetime symmetries as discussed above, but also on the weak field approximation to GR, in the sense that it is in this limit that the EFE becomes a linear equation, and classical plane wave solutions can be found, to which gravitons correspond on quantization. The situation is just as in typical QFTs in which particles are defined for the free, linear equations of motion. (The general picture in QFT was sketched in an appendix to chapter \ref{ch:N-3}: the linearity of the free equations mean that general solutions can be treated as a sum of independent plane waves; on quantization they correspond to particles of definite momentum, and hence general free quantum fields are built from superpositions of many quanta. The specific application to GR is nicely reviewed in \citet[\S2.1]{Kie:12}.) In perturbative, interacting field theory -- including quantum GR -- one assumes that the `Fock space' of arbitrary numbers of such states is a valid approximation, even when interactions, sources, and background fields are present.\footnote{Even though Haag's theorem tells us that it cannot be exact: see \citet[\S10.5]{Dun:12} and \cite{Fra:06}} This is a significant assumption in the case of gravity, since the weak field approximation is not even empirically adequate to the solar system, as the incorrect perihelion of Mercury mentioned above shows. Thus understanding realistic astronomical or cosmological models as graviton coherent states requires the approximate validity of the graviton picture \emph{beyond} the weak field. 

We emphasize here the crucial point that the derivation of the EFE in (D) does \emph{not} depend on the weak field limit, but on the small value of the expansion parameter $\alpha'$ in (\ref{eq:sch2}). We introduced $\alpha'$ as the reciprocal of string tension, but (with $\hbar=c=1$) it has the units of \emph{length}$^2$, so is also equivalent to a fundamental length scale in the theory: $\ell_s\equiv\sqrt{\alpha'}$, the `string length' (familiar from T-duality). Then the rightmost sum in (\ref{eq:sch2}) can be rewritten (as it strictly should be) as an expansion in a \emph{dimensionless} parameter $\ell_s/r$, where $r$ is the only other length scale, the `radius of curvature' of the background spacetime (see \citet[\S3.4.2]{GreSch:87}). Physically then, the approximation that leads to the EFE is valid as long as spacetime is approximately flat over string length sized regions. That is potentially a vastly greater regime of applicability than the weak field regime: the string length is usually taken to be within a few orders of magnitude of the Planck length, so one only expects significant (perturbative) stringy deviations from GR close to the scales at which spacetime has to be treated quantum mechanically.\footnote{This expectation may not always be realized. For instance, the `fuzzballs' proposed for the interior of black holes are not thought to be restricted to a small region around the singularity, but to extend to the horizon.}   Thus  the procedure of taking a background GR model, and computing stringy quantum perturbations around it (to calculate scattering amplitudes, say) is on a solid theoretical footing (notwithstanding its critics' objections). Rather the issue is how the classical metric field emerges in string theory: does the account given in (C) make sense for arbitrary spacetimes? 

Before we address that question, let us briefly discuss what can be said outside the perturbative regime, in situations in which spacetime is curved over regions that are small on the scale of $\ell_s$. First, as we saw in chapter \ref{ch:N-2}, T-duality shows that the classical conception of length is only has operational significance to the string length $\ell_s$; in the familiar sense of the word, `length' is an effective concept, valid only above the string scale. Second, even the notion of target space is predicated on the perturbative approach to string theory; one expands around a supposed classical solution of an unknown exact theory. When $\ell_s/r>1$ the approach breaks down, and one cannot simply assume target space. This is not necessarily to say that perturbative string theory has a non-spacetime phase -- which would realize \cite{Ori:18}'s strongest sense of emergence -- but rather to say that the theory does not apply in such situations. The nature of this regime has of course been subject to considerable scrutiny, because it may offer clues to `M-theory' and because of its cosmological importance. For instance, \cite{Hor:90} raises the question, and \cite{Gre:97} and \citet{Hor:05} explore it using string dualities to look for non-perturbative signs of sub-$\ell_s$ physics. \citet{Gas:07} specifically investigates the possible `string phase' physics of the big bang. There are indeed strong signs that the fundamental degrees of freedom of exact string theory, will not be spatiotemporal, and that there will be states in which spacetime concepts are not well-defined. However, our investigation is into emergence in perturbative string theory, so we will postpone these questions for another occasion, and return to the issue of whether all GR solutions can be understood in terms of dynamical string states.

The situation we have seen thus far is this: the derivation (D) of the EFE is valid without assuming the weak field limit; however, the description (C) of the metric field in terms of coherent graviton states depends on the validity of the graviton picture, which in turn does depend on the weak field approximation. One should not, however, conclude that only in the weak field limit is the metric derived, and otherwise is a given in the fundamental theory. Instead, one should take the success of the graviton picture to indicate that the classical metric should be taken to describe a string state \emph{whatever} the GR solution, but that the description of that state in terms of gravitons is a better or worse approximation depending on how far the spacetime is from the weak field. We know of nothing significant written on the topic; however, in our conversations with string theorists while we have found a range of attitudes about how seriously to take the account given in (C), we have found unanimity that it does not hold in general, but that when it fails, the metric represents some string state not well-described by gravitons. The nature of this state is not understood, and likely requires moving beyond perturbative string theory to the sought-for M-theory.

\subsection{Graviton coherent states}

The central question then becomes whether it is possible to construct classical spacetimes as quantum field states at all. In an appendix to chapter \ref{ch:N-3} we saw: (i) a coherent state corresponds to a Gaussian wavefunction(al) over canonical position (and momentum), with minimal simultaneous uncertainty for both; (ii) a free, linear system will remain coherent, but an interacting, non-linear system will lose coherence.\footnote{Note the confusing terminology: the `cohere' in `coherent state' and `decoherence' does not have the same meaning. Losing coherence in our sense is not the same as decoherence in the sense of approaching a mixed state -- it generally means evolving into a pure state which is not coherent as we have defined it.} One could question whether (i) is sufficient for an approximately classical state. But this is a general question of the interpretation of QM, so we will pass over this issue, and make the standard assumption that it is, because it minimizes quantum uncertainty. (Indeed, we assume that for this reason even approximately coherent states correspond to classical states.) Our focus instead is on (ii). How long do coherent states of non-linear fields remain coherent? In particular, will a coherent state of the gravitational field persist for the duration of a spacetime model, say the life of a universe like ours, so that it can be understood quantum mechanically in terms of  a graviton field?

In the first place, as far as we have been able to determine, from the literature, and from discussions with physicists, although coherent states are generally understood to represent a classical limit of QFT, little or nothing has been proven in full QFT  regarding fields. This is a significant lacuna in the literature (though see \cite{Ros:13}). 

However, in the context of quantum gravity, one can approach the problem using the `superspace' approximation, using a restricted set of degrees of freedom, instead of field quanta. In particular, one can study models of gravity in a `minisuperspace', a finite dimensional configuration space: for instance, for a Friedmann universe described by a scale factor $a(t)$, with a massive scalar field $\phi(t)$. One quantizes by postulating a wavefunction $\psi(a,\phi)$ over this space (see \citet[\S8.1.2]{Kie:12} for more details). From the QFT point of view, one has thus aggregated the microscopic state of field quanta into two degrees of freedom, so a coherent state cannot be given explicitly as an excited state, but rather (as seen in an appendix to chapter \ref{ch:N-3}) will be described by a Gaussian $\psi(a,\phi)$. Just as for a particle, such a state corresponds to a maximally classical quantum state, with equally localized canonical position and momentum, and saturating the uncertainty relations. (From a quantum gravity point of view, one has restricted attention to states which can be described by -- superpositions of -- suitable classical states.)

In that formulation, one can pose the question of the persistence of coherence of the state: how long would it remain (approximately) an uncertainty minimizing Gaussian? The duration of a universe? This question has indeed been studied, with mixed results. On the one hand, \cite{KieLou:99} show, for instance,  that a Sch\-warz\-schild black hole coherent state will take around the Hawking evaporation time to disperse. It follows that over that time period one can rely on the classicality of the stringy graviton background, and hence (ceteris paribus) on results obtained from string theoretic models of black holes. Since such models have been an active area of research, this result is significant. 

On the other hand, \cite{Kie:88} investigates `big crunch' Friedmann universes, which start and end with a singularity. The corresponding quantum state has, as it were, components describing the epochs including each singularity as coherent states, which interfere around the `turnaround' in which the universe starts to recontract, leading to a loss of coherence for that epoch. That is, this quantum model does not allow one to understand the appearance of a full classical spacetime in terms of coherent states. 

Moreover, \citet[\S3.3]{Wal:12} (drawing on \citet{ZurPaz:94,ZurPaz:95,ZurPaz:95a}) argues that coherence will not typically be preserved in realistically complex systems; he considers particle systems, but since these couple to fields, the same should apply to fields as well. Non-linear effects in the classical dynamics lead to chaotic behaviour: initially close states will become observably distinct. But since coherent states track classical states, such behaviour will, on quantization, entail the spreading of wavepackets, and hence the loss of coherence. A relevant example for our purposes, since it involves the gravitational field, is the chaotic motion of Hyperion around Saturn: \cite{ZurPaz:95a} argue that coherence will be observably lost with a few decades.

These studies and arguments are carried out within the study of `decoherence' and the many worlds interpretation. For instance, \citet[\S8.1.2]{Kie:12} and \cite{Kie:13} argue that in minisuperspace models interactions between the aggregate degrees of freedom and the microscopic ones of the quanta leads to decoherence into an approximate mixture of two coherent states. While Wallace is arguing that the coherent state concept alone is not sufficient to recover a classical world. In that context, the papers argue that decoherence will lead the system into an approximate mixture of coherent states. If correct, although coherent states indeed represent classical states for the reasons explained above, one cannot simply identify a realistic, diachronic classical world as a unitarily evolving coherent state. If this general line of interpretation is pursued, in some way different possible classical worlds will have to be identified with decoherent branches -- each consisting of a diachronic series of coherent states. One could follow Wallace, and view each branch as an equal world; one could adopt the view that a particular branch is picked out by `hidden' variables; or perhaps something else. It is not the business of this discussion to adjudicate this question: we have bracketed questions of the interpretation of QM. However, we do point out the scope and ultimate limit of the coherent state concept in representing the classical world in quantum mechanics. We commend this important issue to readers for investigation.

\subsection{GR from QFT}

The final issue concerns the relation of the string theoretic derivation of the metric to the treatment of gravity in QFT, and in particular to derivations of universal coupling (hence weak equivalence) and the (quantized) EFE by Weinberg, in \cite{Wei:64} and \cite{Wei:65}, respectively. Working within the S-matrix approach (see \citet{Cus:90}), Weinberg first shows that Lorentz invariance entails that all \emph{low-energy} massless spin-2 quanta must couple to matter fields with the same coupling strength, the same value for Newton's constant. (The restriction to low energy of course allows for high energy quantum corrections to the S-matrix.) This derivation is examined in a nice paper \cite{Sal:18}, arguing that the need to unify GR and QM follows from relativistic quantum mechanics: (i) from Wigner we have the classification of possible particles, including gravitons; while (ii) from Weinberg we have the necessity of quantum fields (this, along with the other results discussed here, is explained in \cite{Wei:05}), hence of second quantizing gravitons; and then (iii) we have Weinberg's derivation of weak equivalence, describing the graviton field's interactions with other fields; overall, entailing a unified QFT treatment of gravity and matter. (Of course, this unification is not satisfactory, because the resulting field theory is non-renormalizable, as discussed in footnote \ref{ftnt:GRrenorm}.)

In the later paper Weinberg goes further, and working in Dyson-Feynman perturbation theory shows that Lorentz invariance means that any massless spin-2 quantum field will satisfy the quantized EFE as its Heisenberg equation of motion (i.e., with fields replaced with operators); plus possible `Fermi terms'.\footnote{\label{ftnt:GRrenorm}These include couplings to derivatives of curvature, which would violate minimal coupling. Since gravity is non-renormalizable, such terms will have to appear as counterterms at increasing orders of perturbation theory. See \citet[\S2.2]{Kie:12} for a full discussion of this point. (Note that QFT gravity has predictive physical content, despite its non-renormalizability; like any such theory, only a finite number of constants need be empirically fixed at any given perturbative level in order to make renormalized predictions.) If these terms have small couplings at low energy, then the EFE will be the approximate equation of motion, and they will be quantum corrections. (We should also mention that the non-renormalizability of the theory might after all be acceptable, if it is `asymptotically safe': e.g., \citet[\S2.2.5]{Kie:12}.)} (Note that given the Lorentz invariance of the `matter' fields, we have another derivation of Read's two `miracles', this time in QFT.) As explained in an appendix to chapter \ref{ch:N-3}, it follows that the expectation values of the coherent states of these fields will obey the classical EFE, essentially the same result that we found for strings. Does this mean that after all there is nothing so remarkable about finding gravity in string theory?

That conclusion would require accepting that quantum fields are a good description of states of many excited strings. Given the correspondence in symmetry properties, they must be if string theory is to make good physical sense, but that conditional is not trivial. Thus the stringy derivation of the EFEs is an important consistency check on the interpretation of string theory in terms of quantum fields -- they do satisfy the same equations of motion. Moreover, there is the very interesting question of the different sources of the result in QFT and string theory: Lorentz invariance, and worldsheet Weyl symmetry, respectively. (We did see in \S\ref{sec:StringQ} of chapter \ref{ch:N-1} that these symmetries are traded off in lightcone quantization.) Why do a global spacetime symmetry and a local symmetry of the string lead to the same place?\\




\section{Conclusions}

This chapter framed the issue of spacetime emergence in terms of the origin of local and global topologies, and of the metric of spacetime; insofar as these are not baked into the formalism of string theory but derived, spacetime can be said to `emerge' in a broad sense. (After all, their appearance only at the level of classical spacetime theory has considerable novelty.) We emphasize that these questions can be meaningfully addressed without having to address the question of spacetime substantialism, which we view as a methodological virtue.

Duality has been seen to entail that both local and global topology is derived, hence emergent. In the first case, there is no fact of the matter about how observable processes -- specifically scattering events -- are localized in target space. In the latter case, various dualities exist linking global topologies. And we take the stance, developed in chapter \ref{ch:N-2}, that duals are physically equivalent, so that only the `common core', in some sense, has physical significance. Thus the topology is derived.

The metric has been seen to arise from an excited -- graviton -- state of strings. Moreover, we have seen that it is indeed the metric of GR, since it must satisfy the EFE (to lowest order). We have sketched the scope of this derivation, with respect to different target space geometries, and the range of spacetime models that can be described. But the most important point is that the spacetime part of a string solution is arbitrary, and can be traded for gravitons: hence even the partial metric is not fundamental, but conventional -- a choice of perturbative framework. In this sense only the full metric, including graviton field is physical, but it is derived because of the graviton contribution; hence again derived. (These considerations are supported by models, mentioned in passing, of the big bang and black holes in which there is a non-spatiotemporal string phase.)

In other words, these three structures are all derived, hence emergent in the broad sense: not fundamentally present, and of considerable novelty in effective physical theories. However, what of the derivations: what principles do they invoke, and are they `physically salient'? To what extent do they allow us to understand how spacetime might emerge from the non-spatiotemporal? The general principles invoked are: the identification of string excitations with quanta (through the identification of representations of the spacetime symmetries); the use of the tools of QFT (such as coherent states, anomalies, and renormalization); the operationalization of spacetime magnitudes (through scattering amplitudes); and the use of perturbation theory. The first of these seem to have familiar and solid physical salience, but the last is problematic.

In short, we saw that perturbative string theory functions by postulating that the exact (but as yet undiscovered) theory contains classical spacetime solutions, about which the theory expands: in the simplest case, Minkowski spacetime. The worry in the current discussion is not whether such a theory exists (if it does, and is empirically successful, it then must contain classical spacetime models), but rather \emph{what it means} for it to contain a classical spacetime solution. For that is just the question of how to derive the spatial from the non-spatiotemporal, and even though different choices of metric are equivalent, that some metric is introduced entirely begs the question on physical salience: in what way does it arise from some more fundamental structure? Classical spacetime is derived in perturbative string theory, but what it is derived from and how, is left obscure. We will return to this issue in the conclusion.



\newpage

\renewcommand{\thesection}{\Alph{section}}
\renewcommand{\thesubsection}{\Alph{section}.\arabic{subsection}}
\setcounter{section}{0}


\newpage

\bibliographystyle{plainnat}
\bibliography{biblio}

\end{document}